\documentclass[reprint,superscriptaddress,aps,pre]{revtex4-2}

\usepackage{natbib}
\usepackage{graphicx}
\usepackage{amssymb}
\usepackage{amsmath}
\usepackage{epsfig}
\usepackage{chemarr}
\usepackage{url}

\usepackage{dcolumn}
\usepackage{bm}
\usepackage{color}
\usepackage{braket}

\def\be{\begin{equation}}
\def\ee{\end{equation}}
\def\bmu{\begin{multline}}
\def\bea{\begin{eqnarray}}
\def\eea{\end{eqnarray}}

\newcommand{\overbar}[1]{\mkern 1.5mu\overline{\mkern-1.5mu#1\mkern-1.5mu}\mkern 1.5mu}

\setcitestyle{numbers}

\begin{document}

\title{Learning to control active matter}

\author{Martin J Falk}
\affiliation{Department of Physics, University of Chicago, Chicago, IL 60637, United States of America}
\author{Vahid Alizadehyazdi}
\affiliation{Department of Physics, University of Chicago, Chicago, IL 60637, United States of America}
\author{Heinrich Jaeger}
\affiliation{Department of Physics, University of Chicago, Chicago, IL 60637, United States of America}
\author{Arvind Murugan}
\affiliation{Department of Physics, University of Chicago, Chicago, IL 60637, United States of America}

\begin{abstract} 
The study of active matter has revealed novel non-equilibrium collective behaviors, illustrating their potential as a new materials platform.
However, most works treat active matter as unregulated systems with uniform microscopic energy input, which we refer to as activity. In contrast, functionality in biological materials results from regulating and controlling activity locally over space and time, as has only recently become experimentally possible for engineered active matter.
Designing functionality requires navigation of the high dimensional space of spatio-temporal activity patterns, but brute force approaches are unlikely to be successful without system-specific intuition.
Here, we apply reinforcement learning to the task of inducing net transport in a specific direction for a simulated system of Vicsek-like self-propelled disks using a spotlight that increases activity locally. The resulting time-varying patterns of activity learned exploit the distinct physics of the strong and weak coupling regimes. 
Our work shows how reinforcement learning can reveal physically interpretable protocols for controlling collective behavior in non-equilibrium systems.
%
\end{abstract}

\maketitle

\section{Introduction}

Active matter has revealed exciting new patterns of self-organization not found in equilibrium systems\cite{gompper20202020,vicsek1995novel}. Pioneering theoretical and experimental work explored the complexity generated by spatio-temporally uniform systems, in particular with uniform non-equilibrium microscopic driving across space and time. The resulting phenomena are sometimes seen as a step towards achieving complex functionality shown by biological materials.
However, biological functionality, e.g., cytokinesis or cell migration\cite{staddon2018cooperation,cheng2017intermediate,streichan2018global}, can be attributed to not merely being out of equilibrium but rather, to the ability to regulate activity as a function of space and time. 

Recent experimental advances have demonstrated  regulation of activity in diverse engineered systems, including bacteria, colloids and reconstituted cytoskeletal components; while details differ, these experimental platforms allow for activity to be modulated as a function of space and time, usually through optical means\cite{ross2019controlling,zhang2021spatiotemporal,volpe2011microswimmers,buttinoni2012active,palacci2013living,frangipane2018dynamic}.

However, we do not currently have systematic computational frameworks to exploit these new experimental techniques for manipulating active matter. 
The high-dimensional space of spatio-temporal protocols opened up by these experimental advances cannot be explored through brute-force alone. 
Furthermore, activity is a scalar field, and therefore it is not immediately clear how control of this quantity can achieve complex targets like spatial structure or net momentum transfer. 
For example, while a colloid can be induced to self-propel with light, light only controls the scalar speed at which the colloid self-propels, and not the vector direction.
Consequently, previous work relied on system-specific physical intuition\cite{frangipane2018dynamic,stenhammar2016light,das2019active,geiseler2016chemotaxis,geiseler2017taxis,colen2021machine,shankar2019hydrodynamics} or assumed complete knowledge of the underlying dynamical equations\cite{norton2020optimal}.

In contrast, data-driven approaches can be model-free and have shown promise for similar control problems\cite{cichos2020machine} but typically with a few coupled degrees of freedom such as single-particle navigation\cite{haeufle2016external,mano2017optimal,yang2020micro,yang2020efficient,colabrese2018smart,colabrese2017flow,muinos2021reinforcement}, or bio-inspired locomotion\cite{mishra2020coordinated,reddy2016learning,reddy2018glider,novati2019controlled, gazzola2016learning,gazzola2014reinforcement,novati2017synchronisation}. 
These works have established data-driven techniques, in particular reinforcement learning\cite{sutton2018reinforcement}, as a powerful tool for tackling control problems in physics.
However, less attention\cite{durve2020learning,norton2020optimal} has been given to many-body non-equilibrium systems of the kind studied here.


\begin{figure}
\centering
\includegraphics[width=0.4\textwidth]{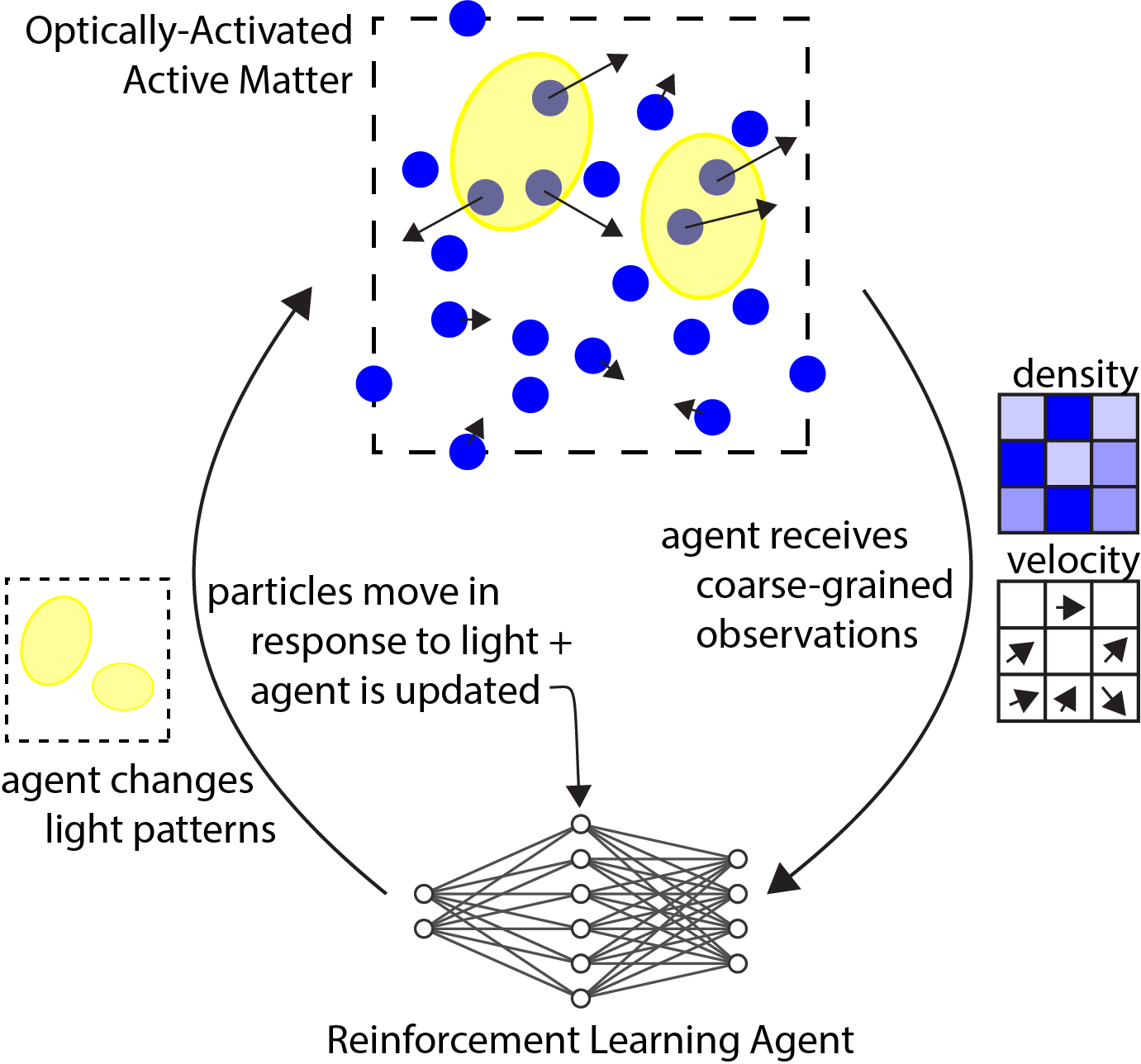}
\caption{\textbf{Reinforcement learning provides a framework to control active matter by modulating activity over space and time.} 
Recent experimental advances allow local modulation of activity by light in diverse active matter systems\cite{ross2019controlling,zhang2021spatiotemporal,frangipane2018dynamic,volpe2011microswimmers,buttinoni2012active,palacci2013living,stenhammar2016light}. We consider a framework in which a reinforcement learning (RL) agent controls the illuminated region - its location, size and shape - in order to achieve a desired non-equilibrium organization. Particle positions and velocity are coarse-grained, and passed to an RL agent, which decides where to place the light. The active matter system responds to the agent's choices of illumination; RL receives a reward based on this response and updates its control protocol accordingly.}
\label{fig:AM_RL_schematic}
\end{figure}



Here, we address the challenge of control in active matter by leveraging developments in model-free reinforcement learning (RL) (Fig. \ref{fig:AM_RL_schematic}).
We construct an RL setup that identifies time-varying patterns of a scalar activity parameter capable of inducing directed transport in a simulated system of Vicsek-like self-propelled disks. As aligning interactions between the disks are increased from zero, the nature of learned protocols changes, illustrating the flexibility of the reinforcement learning approach. We find that the learned protocols can be physically interpreted in terms of the distinct underlying physics at weak and strong coupling. 

In doing so, our goal is to demonstrate that reinforcement learning is a well-suited technique for achieving functionality in a broad class of active systems. 
While the system under consideration here is simple and canonical, it contains two physically very distinct regimes.
The success we demonstrate in each regime is therefore indicative that the performance of the approach is not due to a unique aspect of the regime-specific physics.
Our approach therefore promises to be a useful, model-free tool in confronting the high-dimensional protocol search problem that is universal to optically-activated active matter.

\section{Methods}

\subsection{Simulation Environment Overview}
More concretely, we set out to maximize directional transport in a 2-D system of self-propelled particles by controlling activity.
Particle positions are updated similarly to the canonical Vicsek model\cite{vicsek1995novel,chate2008modeling}, but with the distinction that the magnitude of activity $\nu(\mathbf{x},t)$ is a function of space and time:
\begin{equation}
    \mathbf{x}_{i}(t+\Delta t) = \mathbf{x}_{i}(t) + \nu(\mathbf{x},t)\mathbf{p}_{i}(t) \Delta t  + \frac{1}{\gamma}F_{ex}+ \eta,
\end{equation}
\begin{equation}
    \mathbf{p}_{i}(t+\Delta t) = (U_{\theta} \circ \mathcal{W}) ((1-k)\mathbf{p}_{i}(t) + k\mathbf{\bar{p}}_{i}(t)), 
\end{equation}
where $\mathbf{p}$ is the particle polarization, $\mathbf{\bar{p}}$ its local spatial average, $\Delta t$ a timestep, $U_{\theta}$ a random rotation, $\mathcal{W}$ a normalization, and $k$ a coupling in the interval $[0,1)$, $F_{ex}$ comes from a WCA exluded volume pair-potential, $\gamma$ is a drag coefficient, and $\eta$ is a spatial diffusion term. $\nu(\mathbf{x},t)$ is now a generic spatiotemporal field controlling the speed (but not direction) of active self-propulsion.

As in recent experiments where spatio-temporal control of activity has been achieved optically\cite{ross2019controlling,zhang2021spatiotemporal,frangipane2018dynamic,buttinoni2012active,volpe2011microswimmers,palacci2013living}, we assume that particles' polarities are unaffected by light. 
We stress that the only microscopic quantity changed by the light field is the active propulsion speed. All other quantities, such as rotational diffusion or coupling between polarities, are independent of optical activation. 
As noted, particles also experience excluded volume and a small amount of thermal noise; please refer to Appendix \ref{sim_detail} for more detail on how the simulations are implemented.

In what follows, we will formulate an RL set-up for maximizing the x-component of the system's momentum, in distinct physical regimes.
We emphasize that the goal of maximizing +x-momentum is relatively simple, but requires a non-trivial strategy; as light only controls the scalar speed at which a particle self-propels, not the vector direction, simply illuminating the particles will not produce transport in a specific direction (Sup Movie 1 \cite{SM}).

\subsection{Formulating RL for Active Matter}

Having described the simulation environment, we now turn to the selection of an appropriate algorithm for navigating the space of activity protocols. 
We also have to make choices about how this algorithm will interact with the simulation environment; we need to choose how to represent states, actions, and rewards. 
While most reinforcement learning algorithms are constructed with the goal of enabling successful optimization within a high-dimensional space of protocols, some algorithms will be better suited to the specific requirements of manipulating optically-activated active matter.
We therefore make our choices motivated by the potential application of our approach beyond our simulated Vicsek-like environment.

\subsubsection{Defining states, actions, and rewards}

Reinforcement learning is commonly framed in the language of Markov Decision Processes, which contain four essential ingredients: states, actions, transitions, and rewards\cite{sutton2018reinforcement}. Transitions are determined by the physics of our active matter simulation, but the other three ingredients need to be defined as well.

There are many possibilities for defining states. 
We could, for instance, consider the state of the system to be a list of the positions and velocities of each particle.
However, in order to respect the permutation invariance of the system, we instead construct coarse-grained density and velocity fields (Fig. \ref{fig:AM_RL_schematic}).
While it is possible that the coarse-graining leads to violation of the Markov property for transitions between states, we assume that the grid is fine-grained enough to make these violations quantitatively small.
Furthermore, this approach has the benefit of being easily extended to other common active matter systems which are more readily described on larger length-scales or in terms of fields.

Similarly, there are many possibilities for defining actions, corresponding to the different families of spatio-temporal activation fields.
For simplicity, we constrain our optical field to be a single elliptical light source with fixed intensity, which we term the `spotlight'.
All particles within the spotlight experience the same active propulsion speed. All particles outside of the spotlight are inactive.
The RL algorithm (Fig. \ref{fig:AM_RL_schematic}) is allowed to take actions which change the (a) center, (b) length, and (c) aspect ratio of the spotlight as a function of time, but does not change the intensity or tilt of the spotlight.
We note that, by specifying that there is only one spotlight, we have placed constraints on the space of protocols considered by our RL set-up. 
Any protocols identified in the RL procedure are therefore only guaranteed to be locally optimal in this restricted space.
However, our approach easily generalizes to families of protocols with multiple spotlights that may be necessary to consider in other active matter contexts.

As previously noted, our goal will be for the RL set-up to maximize and maintain the x-component of the system's momentum.
Therefore, we choose to define the reward for any action as the subsequent instantaneous x-momentum in the system.
This is in contrast to other common use cases of reinforcement learning outside of physics, where the reward is frequently sparse in time.





\subsubsection{Selecting an algorithm}

There are a similarly wide array of possibilities for selecting a particular reinforcement learning algorithm.
We wanted to choose one that would be well-suited to active matter systems in general.
As such, we needed to find an RL algorithm that could take advantage of non-sparse rewards, naturally encode stochastic protocols, and accommodate continuous state and actions spaces.


These requirements suggested a class of algorithms known as online actor-critics\cite{sutton2018reinforcement}.
Actor-critics have two components: an actor, and a critic.
The actor is a neural network which receives coarse-grained density and velocity fields, as well as the current size and location of the spotlight. 
Based on this input, the actor samples a change to the pattern of light from a probability distribution, and receives its instantaneous momentum reward.
Aiding the actor in its search is the critic network, which accepts the same state as the actor, but instead outputs an estimate of potential future rewards; in our case, time-integrated x-momentum that can be gained in the future. 
Together, these two networks satisfy the requirements of selecting a reinforcement learning algorithm for an active matter context.


In order to update the two networks, the actor-critic algorithm makes use of the policy gradient theorem\cite{sutton2018reinforcement}, which allows the actor loss function to be written as the product of the log probability of the actor sampling a particular choice for spotlight movement, and the temporal difference error $\delta_{t}$.
$\delta_{t}$ in turn is computed from the critic network, and can be thought of intuitively as the difference between the amount of x-momentum seen in the system following spotlight motion, and how much x-momentum the critic expected to see.
The critic network is then updated with a loss function which quadratically penalizes $\delta_{t}$.
These updates encourage the critic to become more accurate, and encourage the actor to move the spotlight in ways that will outperform the critic's expectations.
Appendix \ref{RL_detail} contains more detail on the actor, critic, states, actions, and rewards.


\begin{figure*}
\centering
\includegraphics[width=0.9\textwidth]{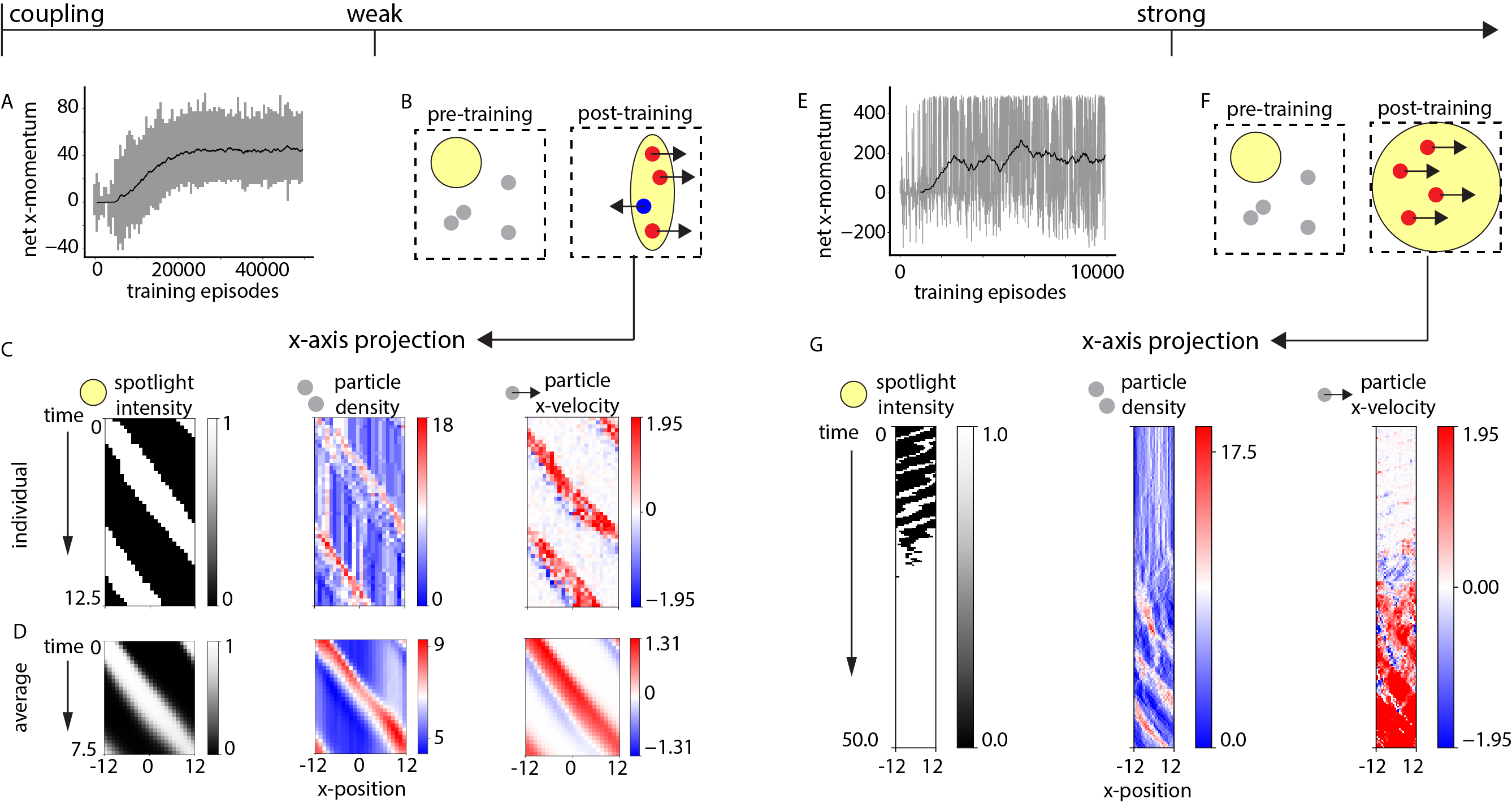}
\caption{\textbf{Reinforcement learning generates distinct protocols to induce directional transport in self-propelled disks at weak and strong coupling.} (weak coupling: \textbf{A}-\textbf{D}, strong coupling: \textbf{E}-\textbf{G}) \textbf{A, E}. As the RL setup trains, average x-momentum during a training episode increases. X-momentum remains highly stochastic (grey curves), but running average of (A)500 or (E)1000 episodes shows clear improvement (black curve). \textbf{B, F}. Training changes spotlight shape and movement in distinctive ways at weak and strong coupling in order to induce rightward transport.  
\textbf{C, G}. Kymographs of spotlight intensity, particle density, and particle x-velocity for the learned protocols in the weak and strong coupling regimes. Kymographs focus on the one spatial dimension important for the target behavior, but some information is lost in the x-projection. Examples of the full system dynamics are available in Sup Movies 2,3 \cite{SM}.
\textbf{D}. For weak coupling, averaging over multiple aligned periods of the learned policy 
shows that the spotlight moves from left to right at a well-defined velocity, with traveling wave of particle density with positive x-velocity carried along.}
\label{fig:fig2}
\end{figure*}

\section{Results}
We begin by exploring control protocols for systems with different coupling $k$ between particle polarities. 
Prior works have established how the physics of Vicsek-like systems qualitatively changes with this parameter\cite{vicsek1995novel,chate2008modeling}, as well as the response of such systems to temporally-fixed quenched disorder of various kinds\cite{toner2018swarming,duan2021breakdown}.
The no-coupling regime has been studied extensively as self-propelled hard-spheres\cite{cates2015motility,reichhardt2017ratchet}. 
As the coupling is increased into a high coupling regime, the system crosses an alignment transition into a flocking phase (see Appendix \ref{vicsek_info}).
In order to achieve +x transport, the RL policy should learn to break the symmetry of the particles' responses and exploit the distinct physics of the two regimes.

In all coupling regimes, the spotlight initially does not move in any meaningful fashion, and the net transport through the system is correspondingly low. As training proceeds, 
net transport through the system increases  (Fig. \ref{fig:fig2}A,E).

In the weak coupling limit, we find that the elliptical spotlight becomes fully elongated in the $y$ direction, of finite length $l_{x}$ in the $x$ direction, and, on average, is moved at a characteristic velocity $v_{\gamma}$ 
in the + $x$ direction (Fig. \ref{fig:fig2}B-D, Sup Movie 2 \cite{SM}). 
A qualitatively distinct strategy with no well-defined spotlight speed and large fluctuations in size is learned at in the strong coupling regime (Fig. \ref{fig:fig2}F,G, Sup Movie 3 \cite{SM}). 



\subsection{Weak Coupling}
We next asked if we could obtain physical insight from our model-free learning approach, starting with the weak coupling limit.

\begin{figure}
\centering
\includegraphics[width=0.45\textwidth]{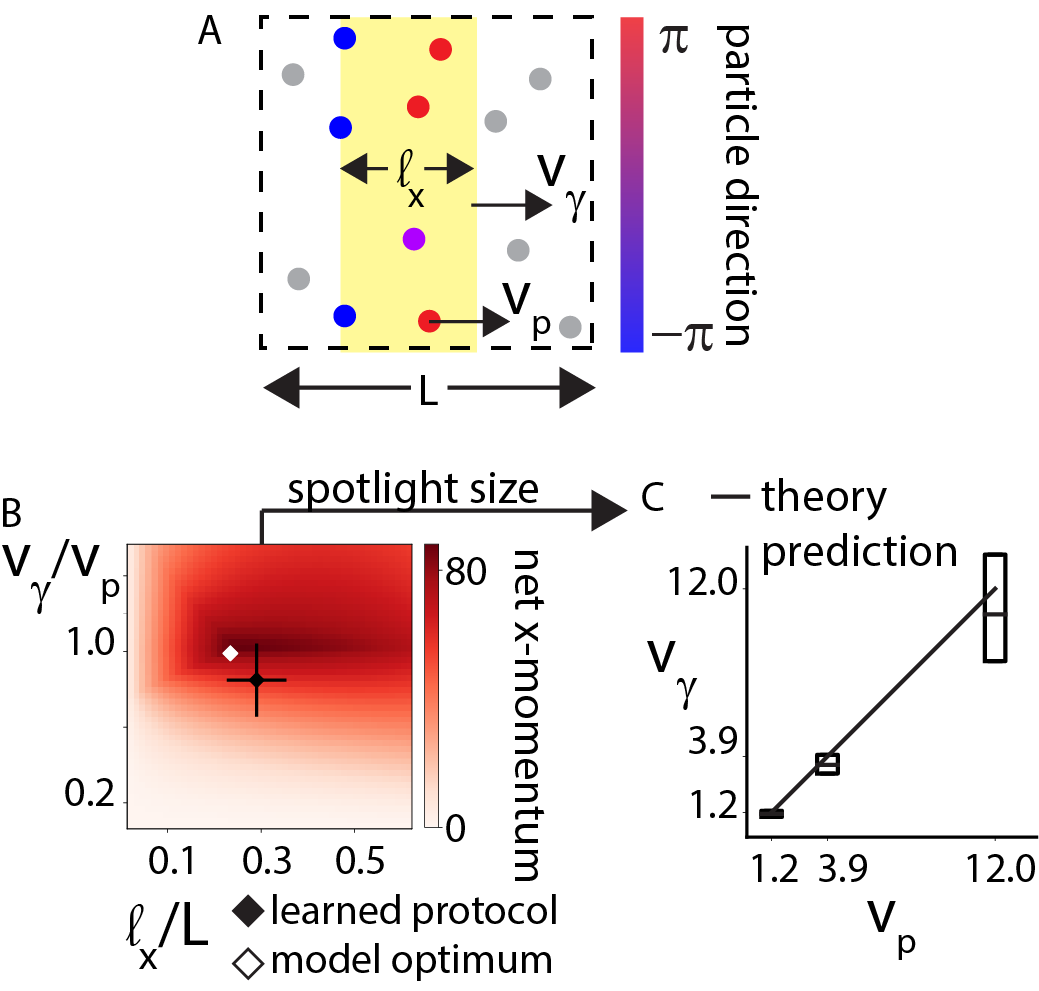}
\caption{\textbf{Weak coupling protocol can be interpreted as a purification process.} 
\textbf{A}. As spotlight moves right, left-moving particles exit the spotlight (and become inactive) sooner than right-moving particles that  co-translate with spotlight. Consequently, -x momentum of left-movers is quenched at the left edge of the spotlight while +x momentum is maintained within the spotlight. At steady-state, density lost by exiting particles is replenished by particles that lie ahead of spotlight's path. 
\textbf{B}. Phase diagram for 1-D model in (A). x-momentum as function of normalized spotlight velocity ($\frac{v_{\gamma}}{v_{p}}$) and normalized spotlight length ($\frac{l_{x}}{L}$). x-momentum is maximized for $v_{\gamma} \approx v_{p}$ and for intermediate $l_{x}$ (black star), close to parameters identified by reinforcement learning (RL)  (white star, black dotted line). Error bars represent the standard deviation of distributions for the trained protocol.
\textbf{C}. We trained RL on self-propelled systems with different light-enhanced activity $v_{p}$. Spotlight speed $v_{\gamma}$ scales with $v_{p}$ as predicted by theory (black line).
Boxplots extend from lower to upper quartile of velocity distribution, with a line at the median. 
}
\label{fig:fig3}
\end{figure}

Based on data in Fig. \ref{fig:fig2}, we propose that the learned policy in the zero-coupling regime functions as a purification process. As the spotlight moves rightward, left-moving particles tend to exit the spotlight quickly, losing activity and reducing -x momentum. In contrast,  right-moving particles tend to remain within the spotlight for longer because both move in the same direction (Fig. \ref{fig:fig3}A), maintaining +x momentum.

We can quantify this intuition using a simplified 1-D model  
in a region of length $L$ with periodic boundary conditions, with an spotlight region of length $l < L$. Particles move with an active speed $v_{p}$ when they are in the spotlight, and there is a conversion rate $r$ of particles that switch their direction of motion per particle per unit time. 

We limit analysis to a protocol where the spotlight moves to the right at a constant velocity $v_{\gamma}$ and make several simplifying assumptions:
(a) number density is a constant $\rho_{a}$ within the spotlight and a constant $\rho_{i}$ outside it.
(b) Particles move with an active speed $v_{p}$ when they are in the spotlight and only experience diffusive motion outside.
(c) The fraction of left-moving particles is a constant $f_{a}$ in the spotlight and $f_{i}$ in the dark.
(d) 
Particles instantaneously randomize their direction of motion upon exiting the spotlight.
(e) The spotlight is a rectangular region of length $l_{x}$ in the +$x$ direction, fully elongated in the $y$ direction. 

These assumptions are broken by real systems and by our simulated system. Furthermore, instances of the learned protocol show deviations that might reflect stochasticity inherent to learning and physical fluctuations such as non-uniform density, finite rotational decoherence time, and other violations of our assumptions.
Nevertheless, we will show that this simple model of purification explains the time-averaged behavior of the learned protocol.


%


\begin{figure}
\centering
\includegraphics[width=0.45\textwidth]{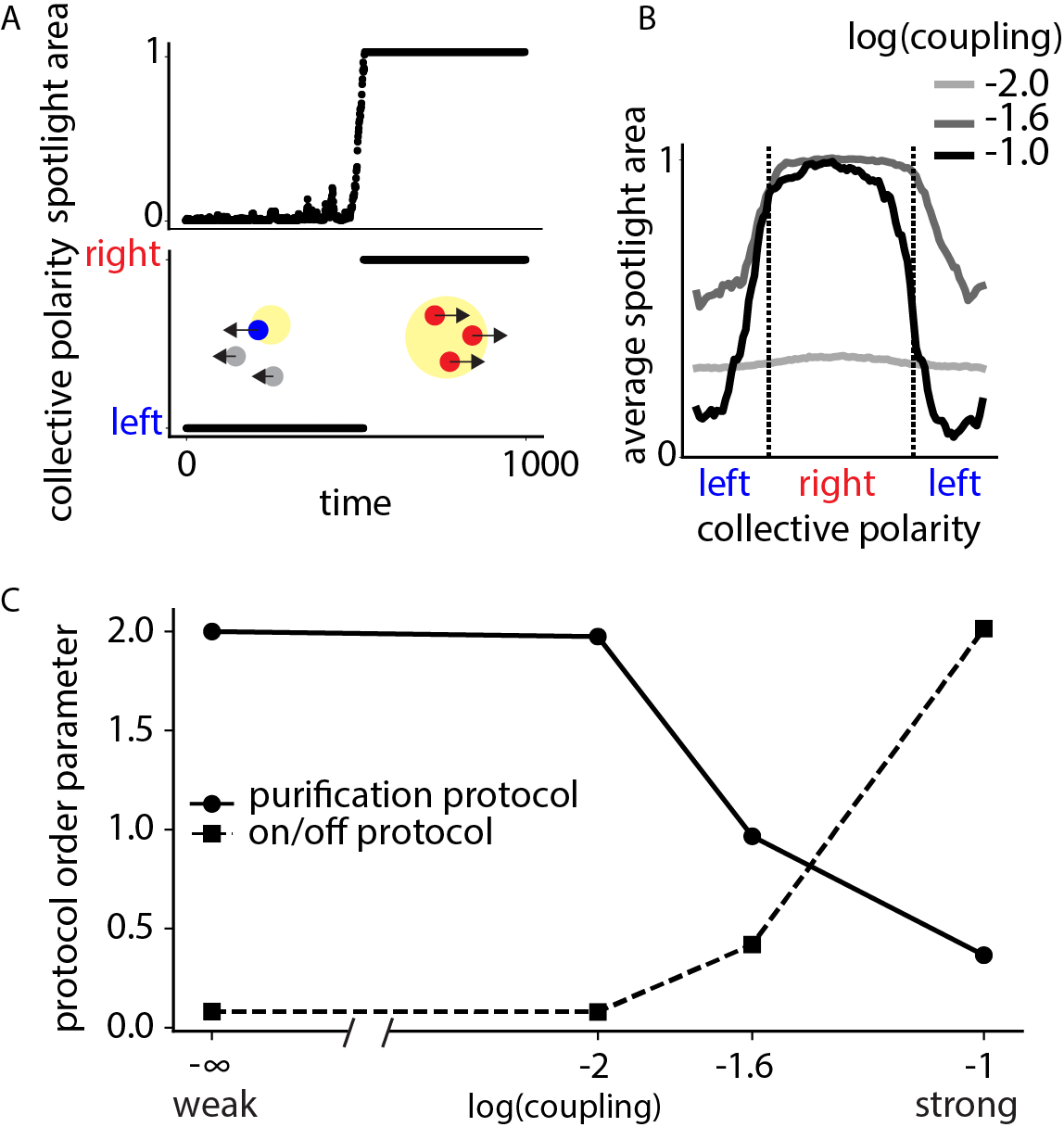}
\caption{\textbf{Order parameters quantify how learned protocols switch from purification at weak coupling to a flocking based strategy at strong coupling.} 
\textbf{A}. Snapshot traces of collective direction (bottom) and ratio between area of spotlight and system area (top) for the strong coupling protocol. Measurements of the two quantities are made at the same time in simulation. \textbf{B}. Average spotlight area as a function of collective polarity. At strong and intermediate coupling, the spotlight provides maximum illumination when the collective polarity points rightward, and provides less illumination when the collective polarity points leftward. There is no relation between collective polarity and spotlight size at weak coupling. This transition is consistent with correlations between spotlight area and collective polarity as a function of coupling (Fig. \ref{fig:sup_fig_4}).
\textbf{C}. Order parameters $\langle P \rangle,\langle O \rangle$ quantitatively track nature of protocols from weak to strong coupling. 
}
\label{fig:fig4}
\end{figure}


To make quantitative connection between the learned protocol and our purification model, we compute +x momentum as a function of purification model parameters.
In the model, we have four unknowns $(\rho_{a},\rho_{i},f_{a},f_{i})$ with four constraints: (a) Particle number conservation
\begin{equation}\label{n_conservation}
\rho_{a}\tilde{l} + \rho_{i}(1-\tilde{l}) = \rho
\end{equation}
where $\rho$ is the overall (linear) number density.
(b) Steady-state particle flux balance into the spotlight
\begin{equation}\label{flux_balance}
(1 + \tilde{v}) f_{a} \rho_{a} + \delta \tilde{v} (1-f_{a}) \rho_{a} = \tilde{v} \rho_{i}
\end{equation}
where $\delta \tilde{v}$ is $| 1 - \tilde{v} |$.
(c) Steady-state flux balance of left-moving particles into the spotlight
\begin{equation}\label{left_a_flux_balance}
(1 + \tilde{v}) f_{a} \rho_{a} + \tilde{r} \tilde{l} \rho_{a} f_{a} = \tilde{v} \rho_{i} f_{i} + \tilde{r} \tilde{l} \rho_{a} (1-f_{a})
\end{equation}
(d) Steady-state flux balance of left-moving particles into the dark region
\begin{equation}\label{left_i_flux_balance}
\tilde{v} \rho_{i} f_{i} + \tilde{r} (1-\tilde{l}) \rho_{i} f_{i} = (1 + \tilde{v}) f_{a} \rho_{a} + \tilde{r} (1-\tilde{l}) \rho_{i} (1-f_{i})
\end{equation}
The equations then involve three non-dimensional quantities, $\tilde{v} = \frac{v_{\gamma}}{v_{p}}$, $\tilde{l} = \frac{l}{L}$, and $\tilde{r} = \frac{rL}{v_{p}}$. To account for the excluded volume of the particles, we modify the system of equations above an adjustable density threshold $\rho_{ev}$.
For further explanation of the various terms in the model, please see Appendix \ref{theory_detail}.


We numerically solve these coupled equations, allowing us to compute a phase diagram for net x-momentum as a function of $\tilde{v}$ and $\tilde{l}$ (Fig. \ref{fig:fig3}B).
We fix $\tilde{r}$ and $\rho_{ev}$ based on values measured in the simulation itself (Appendix \ref{param_estim}).

The phase diagram shows that maximum steady-state momentum is achieved when $v_{\gamma} \approx v_{p}$, consistent with the average velocity of our learned protocol (Fig. \ref{fig:fig3}B) and similar to earlier physics-based single swimmer analyses of periodic activity pulses\cite{geiseler2016chemotaxis,geiseler2017taxis}. Static patterns of diffusivity variations have also been known to generate drift\cite{lanccon2001drift}.

Our theory additionally provides a mechanistic explanation for the existence of an optimal length $l_{x}$ for the spotlight, as we see in our RL-derived policy.
Below this length, the spotlight is too small to accommodate more than a small number of particles, and excluded volume interactions prevent additional accumulation.
Above this length, the spotlight is too big to purify the left-moving particles.
Balancing these two competing effects yields an optimal spotlight length close to the length learned by our RL policy (Fig. \ref{fig:fig3}B).

Finally, the structure of the equations suggests that, so long as we keep $\rho_{ev}$ and $\tilde{r}$ fixed, the velocity of the spotlight $v_{\gamma}$ should be approximately equal to the active speed $v_{p}$. This prediction is confirmed by simulations run at different $v_{p}$ (with $\tilde{l} = .3$) (Fig. \ref{fig:fig3}C).

\subsection{Strong Coupling}
The learned protocol at strong coupling does not have a well-defined spotlight velocity. Instead, spotlight size is correlated with the polarity of particles. 

We find that the strong coupling protocol exploits flocking physics inherent to this regime of the Vicsek model\cite{vicsek1995novel}. Due to the coupling, there is limited heterogeneity in the polarity of individual particles, creating a well-defined collective polarity (Appendix \ref{vicsek_info}). To understand the protocol identified in the strong coupling regime, we compared the collective polarity of the system to the area of the spotlight (Fig. \ref{fig:fig4}A).
We found that RL maximized the area of the spotlight when collective polarity pointed in the desired direction, and fluctuated the spotlight area when 
collective polarity pointed in the undesired direction(Fig. \ref{fig:fig4}B).
This protocol has appealing parallels to other on/off strategies studied, but in the context of single-colloid navigation\cite{haeufle2016external,mano2017optimal,yang2020micro}.

In fact, we can systematically distinguish the learned strategies in the two coupling limits by defining two order parameters that characterize the control protocol. We define the purification parameter $\langle P \rangle$ to be the inverse ratio of the standard deviation and the mean of the $v_{\gamma}$ distribution; intuitively, $\langle P \rangle$ is high if the spotlight has persistent motion in one direction. We define the on/off parameter $\langle O \rangle$ to be the ratio of spotlight area when the collective polarity points right versus left; intuitively, $\langle O \rangle$ is high if the spotlight's intensity is correlated with the collective polarity of particles within the spotlight. 

We repeated the learning algorithm for coupling values between the strong and weak regimes (Fig. \ref{fig:fig4}C). 
We find that $\langle P \rangle$ is high for low coupling and falls with increasing coupling while $\langle O \rangle$ is high for strong coupling and falls with decreasing coupling. Thus, the model-free RL setup learns distinct protocols to exploit different physics at different coupling values.

In the crossover regime ($log(k) = -1.6$) where spontaneous collective motion begins to emerge (Fig. S1, S2), the learned protocol adopts aspects of both the purification and on/off strategies (Fig. \ref{fig:fig4}C).
Like the strong coupling protocol, the agent has maximal spotlight area when the collective polarity points to the right (Fig. \ref{fig:fig4}B).
However, when the collective polarity points to the left, the spotlight area is still half its maximal value on average, potentially reflecting the larger spread of individual particle polarities in the crossover regime. 

Finally, we repeated the learning procedure at strong coupling but lower densities; in this regime, the particles break up into clusters, each with a tightly coupled alignment (Sup Movie 4 \cite{SM}). The learned protocol is harder to directly interpret but achieves a peak momentum transfer closer to the weak coupling regime, despite exhibiting characteristics more similar to the policy learned in the strong coupling regime (Appendix \ref{dilute_regime}, Fig. \ref{fig:sup_fig_3}). 
We additionally repeated the learning procedure at the original density but in larger systems with 4-times the number of particles. In this case, the RL set-up re-identified the same strategies as it did in the smaller systems (Fig. \ref{fig:sup_fig_5}).

\section{Conclusion}
Over the past two decades, the physics of active matter systems with homogeneous activity has been illuminated with great success. Our work proposes that reinforcement learning can be used to explore the collective physics of active particles in spatio-temporally complex environments. 
For the simple models investigated here, we were able to extract physical insight from our initially physics-blind approach, in the strong and weak coupling regimes. Such insight is valuable, particularly given known concerns about the ability of RL to learn reproducible protocols\cite{henderson2018deep}.

Control of active matter by modulating where and when energy is dissipated is broadly applicable since microscopic energy dissipation (i.e., activity) is a universal aspect of active matter systems ranging from bacteria to colloids.
As such, 
we envision that the approach we applied here to particulate active matter can be readily extended to systems where the features of interest are emergent, e.g. topological defects\cite{ross2019controlling,zhang2021spatiotemporal}.
In those systems, non-local effects, e.g. the change in nematic texture due to the motion of topological defects, also suggest the possibility that more complex functional goals will require counter-intuitive solutions.
Similar considerations might also apply to geometrical constraints introduced by confining active matter within deformable containers\cite{quillen2020boids,paoluzzi2016shape,abaurrea2019vesicles,peterson2021vesicle}, or attempting to manipulate an active container filled with passive matter\cite{karimi2020boundary,tanaka2020cable,booth2018omniskins}.
Broadening the nature of the problem, it might also be fruitful to consider the application of reinforcement learning to design the interaction protocols between objects which carry their own sources of illumination\cite{wang2021emergent}.
We therefore propose that reinforcement learning provides an appealing, model-free method for generating intuition and functionalizing the effects of localized activity in systems hosting topological excitations or otherwise complex dynamics.


\textbf{Acknowledgements.} We are indebted to Jonathan Colen, John Devaney, Ryo Hanai, Kabir Husain, Shruti Mishra, Riccardo Ravasio, Matthew Spenko, and Bryan VanSaders for insightful feedback on the manuscript. We would also like to thank Weerapat Pittayakanchit and Steven Redford for helpful discussions at the beginning of the project. This work was primarily supported by NSF EFRI grant 1830939.

\appendix

\section{Simulation Details}\label{sim_detail}

Simulations are run in two dimensional periodic boundary conditions using HOOMD-blue\cite{anderson2020hoomd} (2.9.0). 
Translational dynamics of the system are fairly simple.
A WCA potential between particles is used to enforce excluded volume beginning at a radius of .5. 
Positions are updated under Langevin dynamics, with a drag coefficient of 5, and a small kT of .03. 
Self-propulsion is incorporated into the dynamics via the addition of a constant force whose magnitude is the product of the drag coefficient and the active speed referenced in the text.
Unless otherwise noted, this active speed is set to 3.9.
The direction of the force is updated every five timesteps, and points along the particle's instantaneous polarity at the time of the update.

The angular dynamics of the polarity are also fairly simple. 
Each particle carries with it a unit vector polarity, which points in the plane. 
Every five timesteps, this polarity is rotated by a random angle drawn from the distribution $\frac{\pi}{\sqrt{1000}}\mathcal{N}(0,1)$.

For simulations which incorporate a Vicsek-like coupling between physically proximal particles, every five timesteps the polarity update begins by computing a list of each particle's neighbors which are within a radius of 1.33, including the central particle.
In the following the central particle will be referenced to with the index $i$.
From this list of particles, a mean polarity $\overbar{p_{i}}$ is computed.
The polarity $p_{i}$ of particle $i$ is then updated to be $(1-k) p_{i} + k \overbar{p_{i}}$, where the coupling $k$ can take on values in $[0,1)$.
The updated polarity is then normalized to be of unit length.
As before in the non-interacting case, each polarity is subsequently rotated by a random angle drawn from the distribution $\frac{\pi}{\sqrt{1000}}\mathcal{N}(0,1)$.

In all simulations performed here, 144 particles are initialized on a square lattice with an overall number density of .25 particles per unit area. 
The mass of each particle is set to 1.
Polarities are initialized by drawing from a uniform distribution between 0 and $2\pi$.
Timesteps were set to be $5 \times 10^{-3}$.

\section{Learning Algorithm Details}\label{RL_detail}

We implemented a simple TD actor-critic algorithm\cite{konda2000actor} based on the implementation found in Ref. \cite{Psai2019}, using Tensorflow (2.0.0).

Reinforcement learning is based on the framework of Markov Decision Processes (MDPs), which involve a time-series of states, actions, and rewards.
The formulation of the three essential components of the states, the actions, and the rewards are independent of the specific reinforcement learning algorithm.
We outline those three key-components before briefly describing our implementation of the actor-critic algorithm.

\subsection{States, Actions, Rewards}

States are represented by concatenating the coarse-grained number density field, the coarse-grained x-velocity field, the coarse-grained y-velocity field, and the current position and shape of the spotlight(s).
The fields are all flattened into vectors before concatenation.
In our current study, all fields are $3 \times 3$, so the size of the state space is $3 \times (3 \times 3) + 4 \times (\# spotlights)$, which comes to a total of 31.
The coarse grained velocity fields are constructed by averaging the velocities of all the particles in a particular grid square; if no particles are present, then the velocity is assigned to be zero.
This procedure is accomplished using the SciPy function binned\_statistic\_2d, and the corresponding density field construction is done using the NumPy function histogram2d.
While it is possible that the coarse-graining leads to violation of the Markov property for transitions between states, we assume that the grid is fine-grained enough to make these violations quantitatively small.

The action space is of dimension $4 \times (\# spotlights)$.
Each arm is specified as a 4-tuple: $(x,y,ratio,length)$ where $x$ is the x-position of the center of the arm, $y$ is the y-position of the center of the arm, $ratio$ is the ratio between the extent of the spotlight in the x-direction compared to the y-direction, and $length$ is the extent of the spotlight in the x-direction.
In other words, the region activated by a given arm is an ellipsoid centered at $(x,y)$, with an x-dimension of $length$ and a y-dimension of $\frac{length}{ratio}$.
However, the output of the actor network is not directly these variables, but is instead a vector $(\delta_{1},\delta_{2},\delta_{3},\delta_{4})$.
This effectively regularizes and constrains the policy to be relatively continuous in time.
This vector then updates the current $(x,y,ratio,length)$ to the following values:
\begin{equation}\label{x_update}
x \rightarrow (x+d\delta_{1}+l_{x})\%(2l_{x})-l_{x}
\end{equation}
\begin{equation}\label{y_update}
y \rightarrow (y+d\delta_{2}+l_{y})\%(2l_{y})-l_{y}
\end{equation}
\begin{equation}\label{ratio_update}
ratio \rightarrow \text{clip}(ratio+.1\delta_{3},.1,3)
\end{equation}
\begin{equation}\label{length_update}
length \rightarrow \text{clip}(length+.1\delta_{4},.2l_{x},1.8l_{x})
\end{equation}
where clip is a function that clips the first entry to be within the bounds in the second and third entries, \% is the mod function, and $l_{x}, l_{y}$ are half the widths of the simulation box in the x and y dimensions respectively. 
For the intermediate and strong coupling regimes discussed in Fig. 4, the constraint that $l_{x}$ must be smaller than the full simulation box length is relaxed, and the spotlight is allowed to occupy the full volume of the simulation box:
\begin{equation}\label{length_update_2}
length \rightarrow \text{clip}(length+.1\delta_{4},.2l_{x},2.2l_{x})
\end{equation}

Note that $d$ sets the maximum distance the agent can move the spotlight center between updates.
On physical grounds, this should not be faster than the maximum distance a particle can move between updates. 
Hence, for a given level of activity, we set $d$ so that the spotlight can move 4x the distance that particles can move in the time between spotlight updates.

In order to improve training, we pre-process states before they are fed into the actor and the critic.
Specifically, we generate $10^{5}$ random samples of states the agent is likely to encounter, and then using the sklearn StandardScaler function to define a function which scales states relative to the random samples.
For density fields, this means we sample a 3-by-3 matrix where each entry is drawn from a uniform distribution between 0 and 1, and then normalize and scale this matrix so that the sum is equal to the number of particles in the system.
For velocity fields, we sample two 3-by-3 matrices where each entry is drawn from a uniform distribution between 0 and 1, and then scale them so that entries run between +/- the active speed of the particles.
For the entries corresponding to the spotlight descriptors, we sample four scalars from the [0,1] uniform distribution and scale the first so that it runs between +/- $l_{x}$, the second between +/- $l_{y}$, the third between .1 and 3, and the fourth between $.2l_{x}$ and $1.8l_{x}$. For the intermediate and strong coupling regimes, the fourth was sampled between $.2l_{x}$ and $2.2l_{x}$

The reward is simply the sum of the x-velocities in the system. 

\subsection{TD Actor-Critic}

Given this set-up for states, actions, and rewards, we now describe implementation of a simple TD actor-critic agent; more detail can be found in standard RL references e.g. Ref. \cite{sutton2018reinforcement}.

Our actor-critic agent has two neural networks, the actor and the critic.
We represent the critic as a simple feed-forward neural network with two hidden layers, each with 400 neurons, and a single output node.
The actor is represented as a feed-forward neural network with two hidden layers, each with 40 neurons, and two output layers, each with $4 \times (\# spotlights)$ nodes.
Both networks use ELU activation functions for the hidden layers.
For the actor, one output layer has a linear activation, while the other has a softplus activation.
The one output node for the critic has a linear activation.

During training, the actor receives a state, and its outputs are used to parameterize the means and standard deviations of $4 \times (\# spotlights)$ Gaussian distributions.
These distributions are then sampled and the samples are clipped to be between +/- 1.
These sampled numbers are then turned into actions as described in Appendix \ref{RL_detail}1, the system transitions to its next state according to the physics described in Appendix \ref{sim_detail}, and then a reward is generated based on the next state.

In order to update the two networks, the actor-critic algorithm makes use of the policy gradient theorem\cite{sutton2018reinforcement}, which allows the actor loss function to be written as the product of the log probability of having sampled the action, and the temporal difference error $\delta_{t}$.
$\delta_{t}$ in turn is computed from the critic network, which is a bootstrapped approximation for the difference between the value the critic network assigns to the next state and what value it should actually be.
Since this is a bootstrapped estimate, we approximate what the value should actually be the sum of the reward received in the transition and the value of the previous state.
In order to insure that the critic network outputs will not diverge, the value of the previous state is discounted by a scalar $0 < \gamma < 1$.
The critic network is then updated with a loss function which quadratically penalizes $\delta_{t}$.
All updates are performed just for one learning step, and with a batch size of one, using the Adam optimizer.

In regards to hyperparameters, we attempted to choose standard values which did not need much changing in between the various parameter regimes we considered in this work.
The size of the networks remained the same, and for all networks trained, $\gamma$ is set to $1-10^{-3}$..
The learning rate is $10^{-4}$ for the critic and $2 \times 10^{-6}$ for the actor.
We initially start these values to be 10 times higher, and then quadratically decay the learning rate to their stated values using the tensorflow polynomial\_decay function.

Training is done in episodes of 100 training steps, over which we record the total reward.
Each training step consists of 50 timesteps.
Every 900 episodes, the system restarts to the initial square lattice with randomized polarities.
Fig. 2 agent training was run for $4.95 \times 10^{4}$ episodes, Fig. 3 agents were trained for $6.3 \times 10^{4}$ episodes, and in Fig. 4, the low coupling agent was trained for $2.7 \times 10^{4}$ episodes, while the high and intermediate agents were trained for $9 \times 10^{3}$ episodes.

\section{Theory Description}\label{theory_detail}

Here we provide a more detailed description of Eqs. 3-6, as well as a description of the modifications to account for excluded volume effects.

Eq. 3 is a statement of particle number conservation in our system, which can be written as:
\begin{equation}\label{n_conservation_sup}
\rho_{a}l + \rho_{i}(1-l) = N_{tot}
\end{equation}
The first term on the RHS represents the number of particles in the active region, and the second term represents the number of particles in the inactive region. 
Dividing through by $L$ yields Eq. 3.

Eq. 4 is a statement that at steady-state, the number of particles exiting and entering the active region must be equal:
\begin{equation}\label{flux_balance_sup}
(v_{p} + v_{\gamma}) f_{a} \rho_{a} + | v_{p} - v_{\gamma} | (1-f_{a}) \rho_{a} = v_{\gamma} \rho_{i}
\end{equation}
The RHS first term represents left-moving particles exiting from the back of the active region either because the region has moved past them, or they have propelled themselves into the inactive region.
The second term represents right-moving particles exiting either from the back or the front of the active region due to mismatch between the active velocity and the velocity of the active region.
The LHS accounts for inactive particles moving into the active region as the active region advances. 
Dividing through by $v_{p}$ yields Eq. 4.

Eq. 5 is a statement that at steady-state, the number of left-moving particles exiting and entering the active region must be equal, and this conservation is independent of the balance of total number of particles entering and exiting:
\begin{equation}\label{left_a_flux_balance_sup}
(v_{p} + v_{\gamma}) f_{a} \rho_{a} + r l \rho_{a} f_{a} = v_{\gamma} \rho_{i} f_{i} + r l \rho_{a} (1-f_{a})
\end{equation}
The RHS first term represents left-moving particles exiting from the back of the active region.
The second term represents the loss of the bulk left-moving active population, as formerly left-moving particles flip direction to become right-moving particles.
The LHS terms analogously represent incoming left-moving particles from the inactive region, and addition from particles in the active bulk that switch from right to left.
Dividing through by $v_{p}$ yields Eq. 5.

Finally, Eq. 6 is a statement that at steady-state, the number of left-moving particles exiting and entering the inactive region must be equal:
\begin{equation}\label{left_i_flux_balance_sup}
v_{\gamma} \rho_{i} f_{i} + r (1-l) \rho_{i} f_{i} = (v_{p} + v_{\gamma}) f_{a} \rho_{a} + r (1-l) \rho_{i} (1-f_{i})
\end{equation} 
The physical meaning of the terms is analogous to Eq. \ref{left_a_flux_balance_sup}, and dividing through by $v_{p}$ yields Eq. 6.

In order to account for excluded volume effects, we assume that density within the active region is capped at a value $\rho_{ev}$.
If Eqs. 3-6 initially provide a solution where $\rho_{a} > \rho_{ev}$, then we instead fix $\rho_{a} = \rho_{ev}$.
Note that Eq. 3 still holds, and therefore this implies that both $\rho_{a}$ and $\rho_{i}$ are fixed.
What prevents Eqs. 4-6 from being over-determined is that the physics requires an additional variable $W$ to be introduced in order to account for the particles that are pushed out of the active region as a result of the excluded volume interactions.
In this regime, we solve the following set of equations:
\begin{equation}\label{flux_balance_ev}
(v_{p} + v_{\gamma}) f_{a} \rho_{a} + | v_{p} - v_{\gamma} | (1-f_{a}) \rho_{a} + W = v_{\gamma} \rho_{i}
\end{equation}
\begin{equation}\label{left_a_flux_balance_ev}
(v_{p} + v_{\gamma}) f_{a} \rho_{a} + r l \rho_{a} f_{a} + W f_{a}  = v_{\gamma} \rho_{i} f_{i} + r l \rho_{a} (1-f_{a})
\end{equation}
\begin{equation}\label{left_i_flux_balance_ev}
v_{\gamma} \rho_{i} f_{i} + r (1-l) \rho_{i} f_{i} = (v_{p} + v_{\gamma}) f_{a} \rho_{a} + r (1-l) \rho_{i} (1-f_{i}) + W f_{a}
\end{equation} 
These equations are identical in meaning to Eqs. \ref{flux_balance_sup} - \ref{left_i_flux_balance_sup}, the only difference being the mean-field accounting for the extra population of left-moving particles carried away from the active region by $W$.

\section{Parameter Estimation}\label{param_estim}

In order to compute our numerical phase diagram in Fig. 3B, we need to set two parameters $r$ and $\rho_{ev}$.

In order to estimate $r$, we followed the same process that generates polarity diffusion in our simulation.
We generated 144 random walks of length 100000, where every five timesteps, the value of the walk was changed by a sample drawn from the distribution $\frac{\pi}{\sqrt{1000}}\mathcal{N}(0,1)$.
Particle polarities were initialized uniformly around the unit circle.
Operationally, r is defined as the fraction of particle polarities which switch x-direction in a unit of time.
As the value of the timestep in our simulation is .005 time units, we then downsample our random walk by taking every 200th entry.
We then count the number of times the cosine of the random walks switches sign, and divide that total sum by the number of random walks, and the length of the downsampled walks.
In doing so, we find that for our system, $r = .32$.

In order to estimate $\rho_{ev}$, we make the crude assumption that the $\rho_{ev}$ is the active density $\rho_{a}$ for any optimal protocol, and specifically the optimal protocol found in Fig. 2.
This assumption is based on the intuition that the optimal protocol seeks to maximize $\rho_{a}$ until excluded volume effects saturate the active region, which can be seen in the resultant phase diagram in Fig. 3B.
Measuring this from simulations of the fully converged protocol found in Fig. 2 yields a linear number density of $\rho_{ev} \approx 9$. Since the simulation box is a square of length 24, this corresponds to an area number density of .375.

\section{Aligning interactions in the Vicsek model}\label{vicsek_info}

\begin{figure}
\centering
\includegraphics[width=0.45\textwidth]{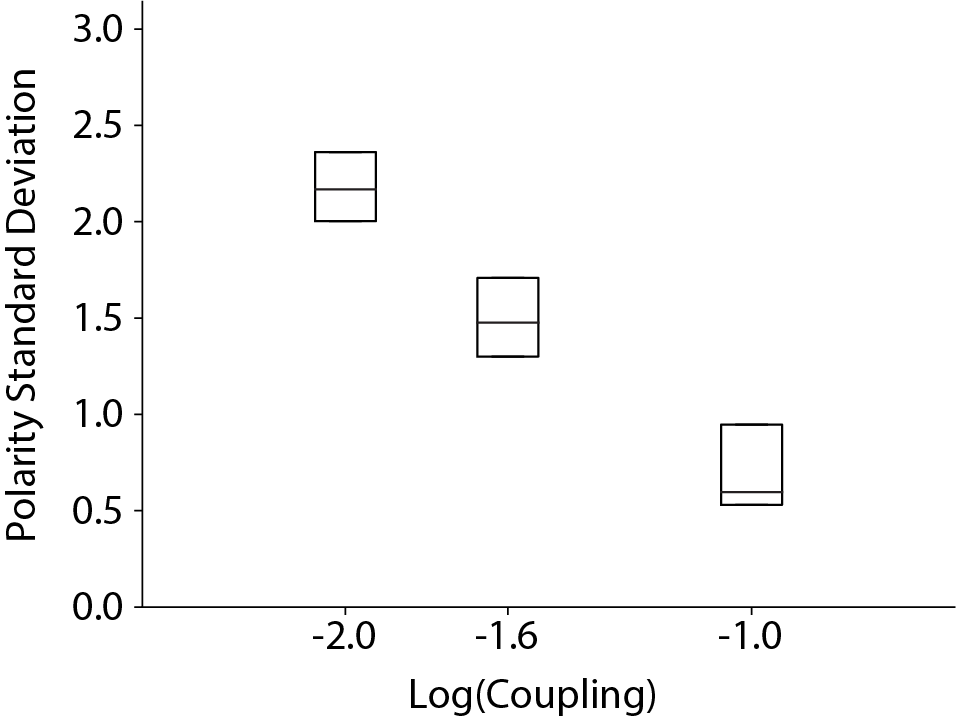}
\caption{\textbf{Increasing coupling sharpens the collective polarity distributions generated in fully-trained trained protocols.} For a range of alignment couplings, we train reinforcement learning agents and evaluate the polarity distributions their converged policies generate. We compute the circular standard deviation at each instance in time and create boxplots to assess the resulting distributions. Each distribution consists of $N = 2.0 \times 10^{5}$ circular standard deviations, drawn from the frames of the same simulations analyzed in Fig. 4. Each circular standard deviation in turn is computed from the polarities of the $n = 144$ particles in the simulation. Boxplots indicate the lower and upper quartiles, with an interior line indicating the median. As coupling increases, the polarity standard deviation decreases, indicating the development of a well-defined collective polarity.}
\label{fig:sup_fig_1}
\end{figure}

\begin{figure}
\centering
\includegraphics[width=0.45\textwidth]{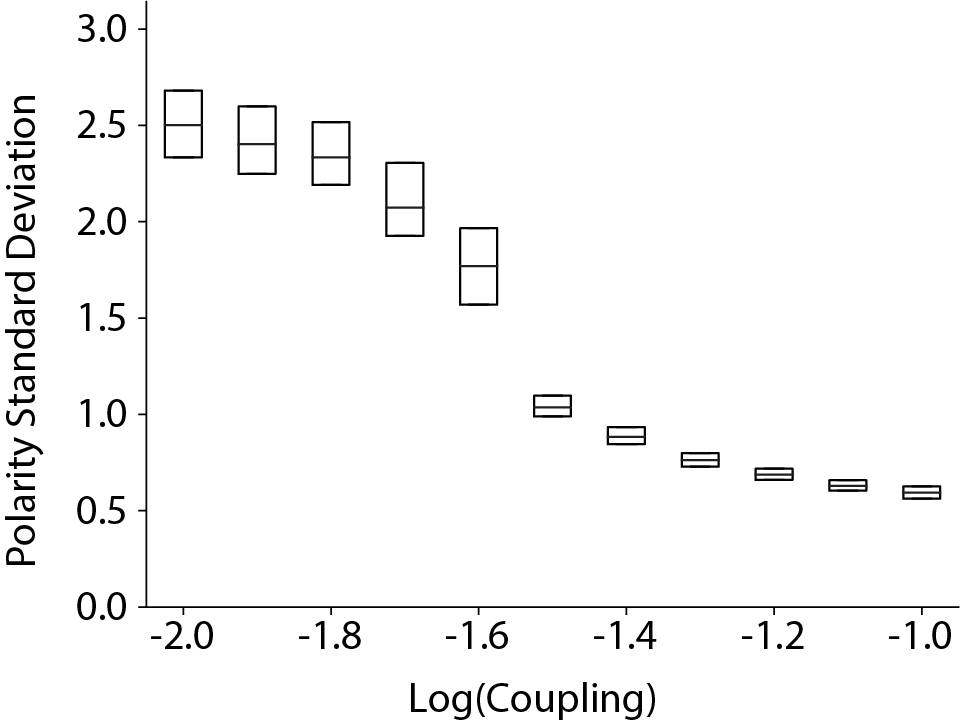}
\caption{\textbf{In a fully-activated Vicsek system, increasing coupling sharpens the collective polarity distribution.} For simulations with permanently activated particles run at a range of polarities, we compute the circular standard deviation at each instance in time and create boxplots to assess the resulting distributions. Each distribution consists of $N = 5.0 \times 10^{3}$ circular standard deviations, drawn from the frames of a simulation where all particles are activated and coupled at the corresponding coupling values. Each circular standard deviation in turn is computed from the polarities of the $n = 144$ particles in the simulation. Boxplots indicate the lower and upper quartiles, with an interior line indicating the median. The system exhibits a crossover regime for polarity standard deviation at a coupling of approximately $k = 10^{-1.5}$, where the standard deviation decreases and the distribution of standard deviations becomes more peaked as well. This indicates the development of a well-defined collective polarity.}
\label{fig:sup_fig_2}
\end{figure}

Central to our analysis of the strong coupling regime is the existence of a well-defined ``collective polarity'', in the sense that the majority of particles have their polarities aligned in the same direction at any given point in time.
That this quantity exists is not surprising considering the well-developed literature on the flocking transition in Vicsek and Vicsek-like models, though the order of the transition is sensitive to computational details\cite{aldana2007phase}.

Here, we can identify on which side of the transition the activated system is using qualitative visual indications (Sup Movie 3 \cite{SM}), and also on a quantitative analysis of the standard deviation of the angular distribution of polarities across time (Fig. \ref{fig:sup_fig_1}).
These distributions are drawn from the same simulations used to generate Fig 4.
We find that as the alignment coupling increases from weak to strong, the distribution of particle polarities has lower average circular standard deviations.
Therefore, even during the periods of inactivity that characterize the agent's policy in the strong coupling regime, the system retains a relatively well-defined collective polarity.

The physics of the Vicsek model do not apply to the inactive periods, and therefore we should not necessarily expect to see a collective flocking polarity during the inactive periods.
However, it is entirely possible that the strength of the coupling and the spatial density of the particles allow for long de-correlation times of polarities of the individual particles, which were aligned during the active periods.
If the decorrelation timescale is long compared to the inactive periods of the policy, then we should still expect to observe a well-defined collective polarity, as we do in the strong coupling regime (Fig. \ref{fig:sup_fig_1}).

To further evidence that the physics of the Vicsek model\cite{vicsek1995novel} underlie the policy learned by the agent in the strong coupling regime, we run simulations identical to those discussed in the main text, except that all particles are activated (Fig. \ref{fig:sup_fig_2}). 
We do this across the range of couplings explored in Fig. 4.
As before in Fig. \ref{fig:sup_fig_1}, we see that the collective polarity becomes more well-defined at stronger couplings.
Additionally, the spread of the distribution of circular standard deviations collected at different time points decreases with coupling.
This decrease indicates that in the strong coupling regime, the majority of particles are aligned the majority of the time, as long as they are constantly activated.

The wider spread at the equivalent coupling values in Fig. \ref{fig:sup_fig_1} are therefore likely to be the result of decorrelation during periods of inactivity. 

Measurements of the circular standard deviation were performed using SciPy's circstd function.

\section{Dilute Regime Policy}\label{dilute_regime}

\begin{figure}
\centering
\includegraphics[width=0.45\textwidth]{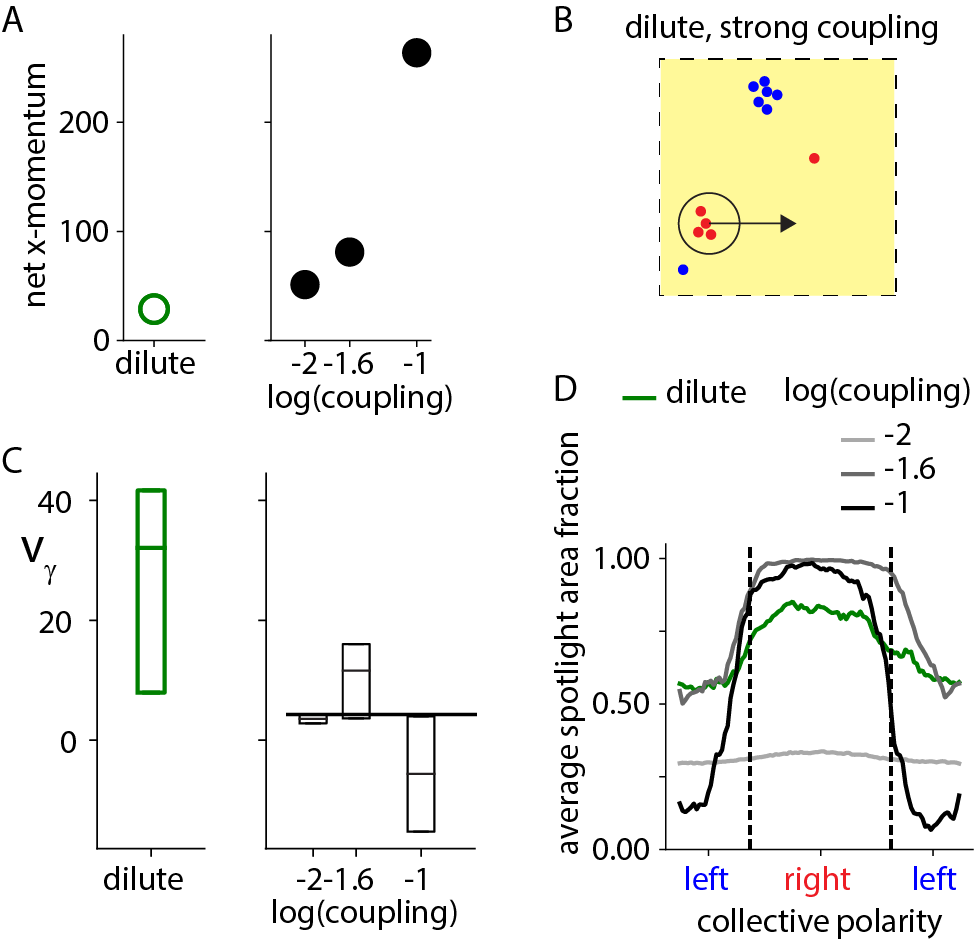}
\caption{\textbf{Learning to induce transport in a strongly coupled dilute system with flocking domains.} All non-dilute data are taken from simulations discussed in Fig. 4. \textbf{A}. In the dilute regime, average  x-momentum transferred by a fully-trained reinforcement learning agent is approximately similar to the amount in the weak coupling regime.
\textbf{B}. Schematic of flocking domains in the dilute, strongly-coupled regime.
\textbf{C}. We evaluate the spotlight x-velocity $v_{\gamma}$ in the dilute regime, as well as over the range of couplings discussed in Fig. 4. At weak coupling, the spotlight moves at the speed predicted by theory (black line). At intermediate and strong coupling, $v_{\gamma}$ diverges from the prediction and additionally is more stochastic. In the dilute regime, $v_{\gamma}$ has an even larger spread.
Boxplots extend from lower to upper quartile of velocity distribution, with a line at the median. 
\textbf{D}. Average spotlight area as a function of collective polarity. In the dilute regime, the spotlight is larger when the average polarity points to the right. 
}
\label{fig:sup_fig_3}
\end{figure}

While Fig. 4 in the main text reports on policies learned for generating transport in systems which exhibited a system-spanning flocking transition, we were also interested in how an RL agent might respond to a system with strong coupling but no system-spanning collective variable.
This is precisely the sort of system which is realized by the Vicsek model when simulated in a dilute regime\cite{vicsek1995novel}.
In the dilute regime, the system-wide flocks break up into flocking domains, with different domain-scale collective polarities (Sup Movie 4 \cite{SM}).

Following training, we find that an RL agent can learn to induce positive x-momentum in a dilute regime (Fig. \ref{fig:sup_fig_3}B), to a degree similar to the weak coupling regime (Fig. \ref{fig:sup_fig_3}A).
Unlike in the weak coupling regime, the spotlight does not move at a well-defined velocity (Fig. \ref{fig:sup_fig_3}C).
The dynamics of the spotlight seems superficially more similar to those trained in the strong coupling regime, with the spotlight larger when the average polarity points in the positive x-direction (Fig. \ref{fig:sup_fig_3}D).

The dilute system has a number density of .033 and an alignment coupling $k = .9$. All other physical constants are the same as those given in Appendix \ref{sim_detail}.
Agents were trained for $1.4 \times 10^{4}$ episodes before training was stopped and the policies were evaluated.

\begin{figure}
\centering
\includegraphics[width=0.45\textwidth]{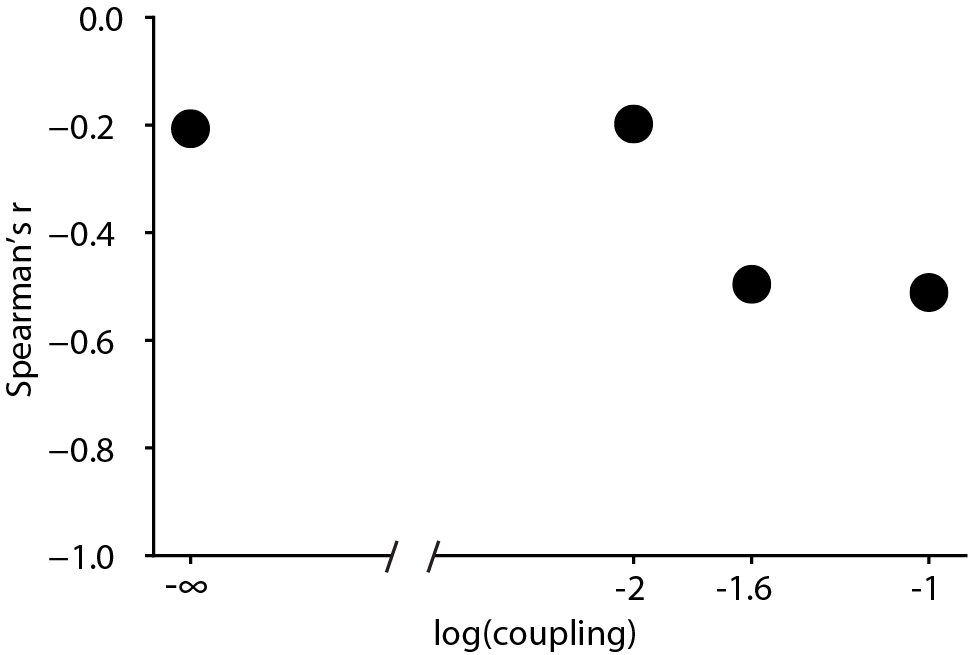}
\caption{\textbf{Spotlight size is anti-correlated with collective motion away from the preferred direction at strong couplings.} We calculate Spearman's r between spotlight area and the absolute value of the angle between collective polarity and the +x direction for increasing values of coupling. As coupling increases, spotlight area and angle magnitude become more strongly anti-correlated.}
\label{fig:sup_fig_4}
\end{figure}

\begin{figure}
\centering
\includegraphics[width=0.45\textwidth]{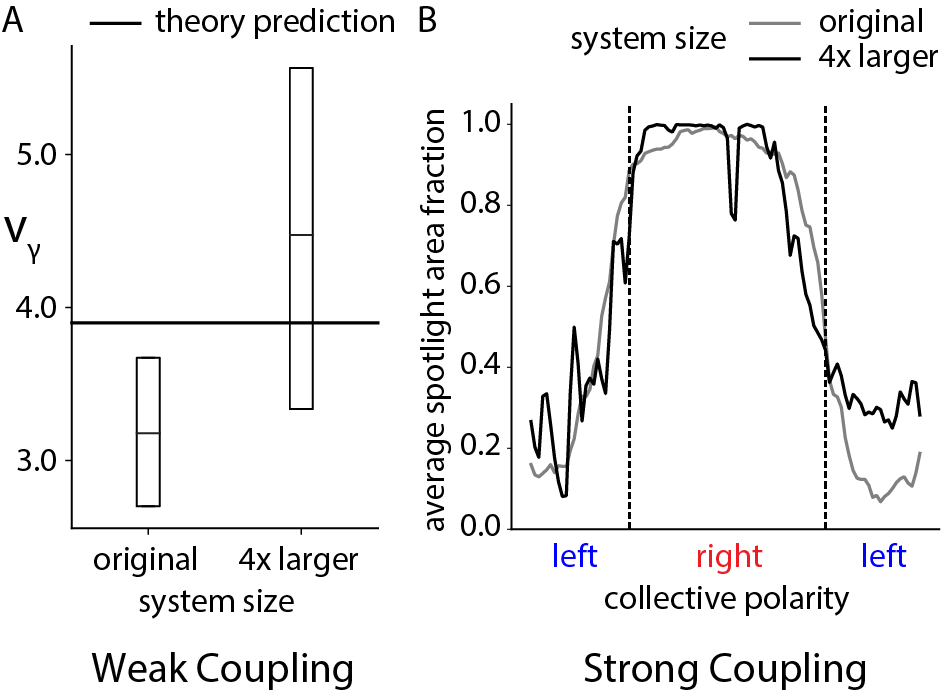}
\caption{\textbf{Learning on 4x larger systems recapitulates the same strategies found in smaller systems for both weak and strong coupling regimes.}
Simulations are run with 576 particles, keeping all other physical parameters constant unless specifically noted. The RL approach and hyperparameters were also identical to those used in the smaller systems.
\textbf{A}. In the weak coupling regime with $k=0$, we find a qualitatively similar strategy to the one identified in Fig. 2, where the spotlight extends laterally and moves rightward at a well-defined velocity, close to the prediction of the simple model proposed in the main text; see also Sup Movie 5 \cite{SM}.
Boxplots extend from lower to upper quartile of velocity distribution, with a line at the median. Training was performed for $1.4 \times 10^{4}$ episodes.
\textbf{B}. In the strong coupling regime with $k=.1$, we find a quantitatively similar strategy to the one identified in Fig. 2, where the spotlight rectifies rightward motion by increasing activity when particles collectively point to the right; see also Sup Movie 6 \cite{SM}. Training was performed for $4.2 \times 10^{3}$ episodes.
}
\label{fig:sup_fig_5}
\end{figure}

\end{document}